\documentclass[twocolumn,showpacs,preprintnumbers,amsmath,amssymb,floatfix]{revtex4}
\usepackage{graphicx}
\usepackage{dcolumn}
\usepackage{bm}

\begin{document}
\title{Neutron-Rich Nuclei in Heaven and Earth}
\author{B.G. Todd-Rutel and J. Piekarewicz}
\affiliation{Department of Physics, Florida State 
             University, Tallahassee, FL 32306}
\date{\today} 

\begin{abstract}
An accurately calibrated relativistic parametrization is introduced to
compute the ground state properties of finite nuclei, their linear
response, and the structure of neutron stars. While similar in spirit
to the successful NL3 parameter set, it produces an equation of state
that is considerably softer --- both for symmetric nuclear matter and
for the symmetry energy. This softening appears to be required for an
accurate description of several collective modes having different
neutron-to-proton ratios. Among the predictions of this model are a
symmetric nuclear-matter incompressibility of $K\!=\!230$~MeV and a
neutron skin thickness in ${}^{208}$Pb of $R_{n}-R_{p}\!=\!0.21$~fm.
Further, the impact of such a softening on the properties of neutron
stars is as follows: the model predicts a limiting neutron star mass
of $M_{\rm max}\!=\!1.72 M_{\odot}$, a radius of $R\!=\!12.66$~km for
a ``canonical'' $M\!=\!1.4 M_{\odot}$ neutron star, and no (nucleon)
direct Urca cooling in neutrons stars with masses below 
$M\!=\!1.3 M_{\odot}$.
\end{abstract}
\maketitle 

The quest for the equation of state (EOS) of neutron-rich matter ---
which is likely to lead to the discovery of exotic phases of matter
--- is an exciting problem that permeates over many areas of physics.
While the search for novel phenomena has long been at the forefront of
science, learning about ``neutron-rich nuclei in heaven and earth''
has experienced a recent revitalization due to remarkable advances in
both terrestrial experiments and space observations. These
developments, coupled to the promise of new {\it facilities for the
future of science}, guarantee continuing discoveries for many years to
come.  Figuring prominently among the facilities for the future is 
the Rare Isotope Accelerator (RIA), a facility that by
defining the limits of nuclear existence, will constrain the EOS at
large neutron-proton asymmetries. In addition, new telescopes
operating at a variety of wavelengths have turned neutron stars from
theoretical curiosities into powerful diagnostic tools.  For some
recent excellent reviews on the relevance of the EOS on a variety of
phenomena, such as the dynamics of heavy-ion collisions, the structure
of neutron stars, and the simulation of core-collapse supernova, see
Refs.~\cite{Dan02_Science298,Lat04_Sci304,Ste04_xxx,Bur03_PRL90} and
references contained therein.

Our aim in this letter is to construct an accurately calibrated
parameter set that, while constrained only by the ground state
properties and the linear response of a variety of nuclei, may still
be used to predict some neutron-star observables. Such a successful
paradigm is the relativistic NL3 parameter set of Lalazissis, Konig,
and Ring~\cite{Lal97_PRC55}. The NL3 parametrization has been used
with enormous success in the description of a variety of ground-state
properties of spherical, deformed, and exotic nuclei. For some special
cases, it has also been used successfully to compute the linear
response of the mean-field ground state. In the particular case of the
giant monopole resonance (GMR) in ${}^{208}$Pb --- the so-called {\it
breathing mode} --- the predicted distribution of strength is in close
agreement with the experimental data~\cite{You99_PRL82}. Thus, it has
come as a surprise that to reproduce the GMR in $^{208}$Pb, accurately
fit nonrelativistic and relativistic models predict compressional
moduli in symmetric nuclear matter ($K$) that differ by about 25\%.
Indeed, while nonrelativistic models predict
$K\!\simeq\!220\!-\!235$~MeV~\cite{Col92_PLB276,Bla95_NPA591,Ham97_PRC56},
relativistic models argue for a significantly larger value
$K\!\simeq\!250\!-\!270$~MeV~\cite{Lal97_PRC55,Vre00_PLB487,Vre03_PRC68}.

In an earlier work, the density dependence of the symmetry energy,
which at present is poorly known, has been proposed as the culprit for
the above discrepancy~\cite{Pie02_PRC66}. Since first proposed, other
groups have tested this assertion reaching similar
conclusions~\cite{Vre03_PRC68,Agr03_PRC68,Col03_NuclTh,Pie04_PRC69}.
In particular, in Refs.~\cite{Pie02_PRC66,Pie04_PRC69} it has been
argued that a good description of the breathing mode in $^{208}$Pb may
be obtained using a large value of $K$ --- provided one compensates
with an appropriately stiff symmetry energy.  Thus, it was suggested
that $^{90}$Zr, a nucleus with both a well-developed breathing mode
and a small neutron-proton asymmetry, should be used (rather than
$^{208}$Pb) to constrain $K$.  With the advent of unprecedented
experimental accuracy in the determination of the breathing mode in
$^{90}$Zr~\cite{You99_PRL82}, it now appears that the NL3 interaction
overestimates the value of $K$~\cite{Agr03_PRC68}.  Moreover, the
alluded correlation between $K$ and the density dependence of the
symmetry energy serves as a telltale of a further problem with most
relativistic parameterizations: an underestimation of the frequency of
oscillations of neutrons against protons --- the so-called isovector
giant dipole resonance (IVGDR) --- in
$^{208}$Pb~\cite{Vre03_PRC68,Pie04_PRC69}.

In this Letter we introduce a new, accurately calibrated relativistic
parametrization that simultaneously describes the GMR in $^{90}$Zr and
$^{208}$Pb, and the IVGDR in $^{208}$Pb, without compromising the
success in reproducing ground-state observables. To do so, however,
two additional coupling constants must be introduced. Without these
additional coupling constants one can not describe the various modes
without seriously compromising the quality of the
fit~\cite{Vre99_NPA649,Vre03_PRC68}.  The effective field theoretical
model is based on an interacting Lagrangian that provides an accurate
description of finite nuclei and a Lorentz covariant extrapolation for
the equation of state of dense matter. It has the following
form~\cite{Mue96_NPA606,Hor01_PRL86,Hor01_PRC64}:
\begin{widetext}
\begin{eqnarray}
&&
{\cal L}_{\rm int} =
\bar\psi \left[g_{\rm s}\phi   \!-\! 
         \left(g_{\rm v}V_\mu  \!+\!
    \frac{g_{\rho}}{2}{\mbox{\boldmath $\tau$}}\cdot{\bf b}_{\mu} 
                               \!+\!    
    \frac{e}{2}(1\!+\!\tau_{3})A_{\mu}\right)\gamma^{\mu}
         \right]\psi \nonumber \\
                   && - 
    \frac{\kappa}{3!} (g_{\rm s}\phi)^3 \!-\!
    \frac{\lambda}{4!}(g_{\rm s}\phi)^4 \!+\!
    \frac{\zeta}{4!}   
    \Big(g_{\rm v}^2 V_{\mu}V^\mu\Big)^2 \!+\!
    \Lambda_{\rm v}
    \Big(g_{\rho}^{2}\,{\bf b}_{\mu}\cdot{\bf b}^{\mu}\Big)
    \Big(g_{\rm v}^2V_{\mu}V^\mu\Big) \;.
\end{eqnarray}
\end{widetext}
The Lagrangian density includes Yukawa couplings of the nucleon field
to various meson fields. It includes an isoscalar-scalar $\phi$ meson
field and three vector fields: an isoscalar $V^{\mu}$, an isovector 
${\bf b}^{\mu}$, and the photon $A^{\mu}$. In addition to the Yukawa
couplings, the Lagrangian is supplemented by four nonlinear meson
interactions. The inclusion of isoscalar meson self-interactions (via
$\kappa$, $\lambda$, and, $\zeta$) are used to soften the equation of
state of symmetric nuclear matter, while the mixed isoscalar-isovector
coupling ($\Lambda_{\rm v}$) modifies the density dependence of the
symmetry energy. While power counting suggests that other local meson
terms may be equally important~\cite{Mue96_NPA606}, their
phenomenological impact has been documented to be
small~\cite{Mue96_NPA606,Hor01_PRL86,Hor01_PRC64}, so they will be not
be considered any further in this study.

\subsection{Ground-state Properties}
\label{groundstate}

Although in Refs~\cite{Hor01_PRL86,Hor01_PRC64} it has been proven
that the addition of the isoscalar-isovector coupling ($\Lambda_{\rm
v}$) is important for the softening of the symmetry energy, no attempt
was made to optimize the various parameter sets. Following standard
practices~\cite{Lal97_PRC55,Agr03_PRC68}, we use binding energies and
charge radii for a variety of magic nuclei to produce an accurately
calibrated set. We dubbed this set ``FSUGold''. Details about the
calibration procedure will be presented in a forthcoming
publication. For now, we present in Table~\ref{Table1} a comparison
between the very successful NL3 parametrization~\cite{Lal97_PRC55},
FSUGold, and, (when available) experimental data. While the agreement
with experiment (at the 1\% level or better) is satisfactory --- and
this agreement extends all over the periodic table~\cite{Lal99_NDT71}
--- a question immediately arises: given the success of NL3, why the 
need for another effective interaction having two additional parameters?
The answer to this question is provided in the next section.
  \begin{table}
  \begin{tabular}{|c|c|c|c|c|}
    \hline
    Nucleus & Observable & Experiment & NL3 & FSUGold \\
    \hline
    \hline
    ${}^{40}$Ca & $B/A$~(MeV)            & $8.55$ & $\phantom{-}8.54$ & $\phantom{-}8.54$  \\
                & $R_{\rm ch}$~(fm)      & $3.45$ & $\phantom{-}3.46$ & $\phantom{-}3.42$  \\
                & $R_{n}\!-\!R_{p}$~(fm) &   ---  & $-0.05$           & $-0.05$            \\
    \hline
    ${}^{48}$Ca & $B/A$~(MeV)            & $8.67$ & $\phantom{-}8.64$ & $\phantom{-}8.58$  \\
                & $R_{\rm ch}$~(fm)      & $3.45$ & $\phantom{-}3.46$ & $\phantom{-}3.45$  \\
                & $R_{n}\!-\!R_{p}$~(fm) &   ---  & $\phantom{-}0.23$ & $\phantom{-}0.20$  \\
    \hline
    ${}^{90}$Zr & $B/A$~(MeV)            & $8.71$ & $\phantom{-}8.69$ & $\phantom{-}8.68$  \\
                & $R_{\rm ch}$~(fm)      & $4.26$ & $\phantom{-}4.26$ & $\phantom{-}4.25$  \\
                & $R_{n}\!-\!R_{p}$~(fm) &   ---  & $\phantom{-}0.11$ & $\phantom{-}0.09$  \\
    \hline
   ${}^{116}$Sn & $B/A$~(MeV)            & $8.52$ & $\phantom{-}8.48$ & $\phantom{-}8.50$  \\
                & $R_{\rm ch}$~(fm)      & $4.63$ & $\phantom{-}4.60$ & $\phantom{-}4.60$  \\
                & $R_{n}\!-\!R_{p}$~(fm) &   ---  & $\phantom{-}0.17$ & $\phantom{-}0.13$  \\
    \hline
   ${}^{132}$Sn & $B/A$~(MeV)            & $8.36$ & $\phantom{-}8.37$ & $\phantom{-}8.34$  \\
                & $R_{\rm ch}$~(fm)      &   ---  & $\phantom{-}4.70$ & $\phantom{-}4.71$  \\
                & $R_{n}\!-\!R_{p}$~(fm) &   ---  & $\phantom{-}0.35$ & $\phantom{-}0.27$  \\
    \hline
   ${}^{208}$Pb & $B/A$~(MeV)            & $7.87$ & $\phantom{-}7.88$ & $\phantom{-}7.89$  \\
                & $R_{\rm ch}$~(fm)      & $5.50$ & $\phantom{-}5.51$ & $\phantom{-}5.52$  \\
                & $R_{n}\!-\!R_{p}$~(fm) &   ---  & $\phantom{-}0.28$ & $\phantom{-}0.21$  \\
    \hline
  \end{tabular}
 \caption{Experimental data for the binding energy per nucleon and the
          charge radii for the magic nuclei used in the least square
          fitting procedure. In addition, predictions are displayed
          for the neutron skin of these nuclei.}
  \label{Table1}
 \end{table}

\subsection{Nuclear Collective Modes}
\label{collectivemodes}

As alluded earlier, and argued in
Refs.~\cite{Pie02_PRC66,Pie04_PRC69}, the success of the NL3 set in
reproducing the breathing mode in ${}^{208}$Pb is accidental; it
results from a combination of both a stiff equation of state for
symmetric nuclear matter and a stiff symmetry energy. If true, this
implies that NL3 should overestimate the location of the breathing
mode in ${}^{90}$Zr --- a nucleus with a well-developed GMR strength
but rather insensitive to the symmetry energy. Similarly, the energy
of the IVGDR in ${}^{208}$Pb, an observable sensitive to the density
dependence of the symmetry energy, should be underestimated by NL3
(note that a stiff symmetry energy predicts a small symmetry energy at
the low densities of relevance to the IVGDR). In
Table~\ref{Table2} relativistic random phase approximation (RPA)
results for the GMR (centroids) in ${}^{208}$Pb and ${}^{90}$Zr, and
the IVGDR (peak energy) in ${}^{208}$Pb are reported. These
small-amplitude modes represent the linear response of the mean field
ground state to a variety of probes~\cite{Rit93_PRL70,You99_PRL82}.
Note that the FSUGold(NL3) parameter set predicts a compression
modulus for symmetric nuclear matter of $K\!=\!230(271)$~MeV and a
neutron skin in ${}^{208}$Pb of
$R_{n}\!-\!R_{p}\!=\!0.21(0.28)$~fm. The good agreement between
FSUGold and experiment is due to the addition of the two extra
parameters ($\zeta$ to reduce the value of $K$ and $\Lambda_{\rm v}$
to soften the symmetry energy). However, it seems that a further
softening of the symmetry energy could further improve the agreement
with experiment. With the present parametrization this could not be
achieved without compromising the quality of the fit. Thus, our
prediction of $R_{n}\!-\!R_{p}\!=\!0.21$~fm could be regarded as an
upper bound.  We note that the Parity Radius Experiment (PREX) at the
Jefferson Laboratory is scheduled to measure the neutron radius of
$^{208}$Pb accurately (to within $0.05$~fm) and model independently
via parity-violating electron
scattering~\cite{Hor01_PRC63,Mic02_JLAB003}.  This experiment should
provide a unique observational constraint on the density dependence of
the symmetry energy.

  \begin{table}
  \begin{tabular}{|c|c|c|c|c|}
    \hline
    Nucleus & Observable & Experiment & NL3 & FSUGold \\
    \hline
    \hline
    ${}^{208}$Pb & GMR   (MeV) & $14.17\pm0.28$ & $14.32$ & $14.04$  \\
    ${}^{90}$Zr  & GMR   (MeV) & $17.89\pm0.20$ & $18.62$ & $17.98$  \\
    ${}^{208}$Pb & IVGDR (MeV) & $13.30\pm0.10$ & $12.70$ & $13.07$  \\
    \hline
  \end{tabular}
 \caption{Centroid energies for the breathing mode in ${}^{208}$Pb and
          ${}^{90}$Zr, and the peak energy for the IVGDR in
          ${}^{208}$Pb.  Experimental data are extracted from
          Refs.~\cite{You99_PRL82} and~\cite{Rit93_PRL70}.}
  \label{Table2}
 \end{table}

\subsection{Neutron Star Observables}
\label{neutronstars}

Having constructed a new accurately calibrated parameter set,
we now examine predictions for a few neutron-star properties. The
structure of spherical neutron stars in hydrostatic equilibrium is
solely determined by the EOS of neutron-rich matter in beta
equilibrium. For the uniform liquid phase we assume an EOS for matter
in beta equilibrium that is composed of neutrons, protons, electrons,
and muons. Further, we assume that this description remains valid in
the high-density interior of the star. Thus, transitions to exotic
phases are not considered here.

However, at the lower densities of the inner crust the uniform system
becomes unstable against density fluctuations. In this nonuniform
region the system may consists of a variety of complex structures,
collectively known as 
{\it nuclear pasta}~\cite{Rav83_PRL50,Has84_PTP71}. While microscopic 
calculations of the nuclear pasta are becoming
available~\cite{Wat05_xxx,Hor04_PRC69,Hor04_PRC70}, it is premature to
incorporate them in our calculation. Hence, following the procedure
adopted in Ref.~\cite{Car03_APJ593}, a simple polytropic equation of
state is used to interpolate from the outer crust~\cite{Bay71_ApJ170}
to the uniform liquid.

Results for the transition density from uniform to nonuniform neutron
rich matter are displayed in Table~\ref{Table3}. These results are
consistent with the inverse correlation between the neutron-skin and
the transition density found in Ref~\cite{Hor01_PRL86}. This
correlation suggests that models with a stiff equation of state (like
NL3) predict a low transition density, as it is energetically
unfavorable to separate nuclear matter into regions of high and low
densities.
  \begin{table}
  \begin{tabular}{|c|c|c|}
    \hline
     Observable & NL3 & FSUGold \\
    \hline
    \hline
     $\rho_{c}$~(fm$^{-3}$)          & $0.052$ & $0.076$  \\
     $R$~(km)                        & $15.05$ & $12.66$  \\
     $M_{\rm max}(M_{\odot})$        & $2.78$ & $1.72$    \\
     $\rho_{_{\rm Urca}}$~(fm$^{-3}$)& $0.21$ & $0.47$    \\
     $M_{\rm Urca}(M_{\odot})$       & $0.84$ & $1.30$    \\
     $\Delta M_{\rm Urca}$           & $0.38$ & $0.06$    \\
    \hline
  \end{tabular}
 \caption{Predictions for a few neutron-star observables. The various
          quantities are as follows: $\rho_{c}$ is the transition
          density from nonuniform to uniform neutron-rich matter
          matter, $R$ is the radius of a 1.4 solar-mass neutron star,
          $M_{\rm max}$ is the limiting mass, $\rho_{_{\rm Urca}}$ is 
	  the threshold density for the direct Urca process, $M_{\rm Urca}$ 
          is the minimum mass neutron star that may cool down by the 
          direct Urca process, and $\Delta M_{\rm Urca}$ is the mass 
          fraction of a 1.4 solar-mass neutron star that supports 
          enhanced cooling by the direct Urca process.}
  \label{Table3}
 \end{table}
We now present results for a few neutron-star observables that depend
critically on the equation of state~\cite{Lat01_APJ550,Lat04_Sci304},
namely, masses, radii, and composition. Table~\ref{Table3} includes
predictions for the radius of a ``canonical'' $1.4$ solar-mass neutron
star alongside the maximum mass that the EOS can support against
gravitational collapse; beyond this value the star collapses into a
black hole. These results were obtained by numerically integrating the
Tolman-Oppenheimer-Volkoff equations. The considerable smaller radius
predicted by the FSUGold model originates in its softer symmetry
energy. The stiffer symmetry energy of the NL3 set does not tolerate
large central densities and produces stars with large radii. Note that
the same physics that pushes neutrons out against surface tension in
the nucleus of ${}^{208}$Pb is also responsible for pushing neutrons
out in a neutron star~\cite{Hor01_PRL86,Hor01_PRC64}.  Further, while
the sizable reduction in the limiting mass of FSUGold relative to NL3
is also due to the softening of the EOS, it is the softening of the
EOS of {\it symmetric nuclear matter} --- rather than the softening of
the symmetry energy --- that is responsible for this effect.

We conclude with a comment on the enhanced cooling of neutrons
stars. Recent observations by the Chandra and XMM-Newton observatories
suggest that some neutron stars may cool rapidly, suggesting perhaps
the need for some exotic component, such as condensates or color
superconductors.  Here we explore a more conservative alternative,
namely, that of enhanced cooling of neutron stars by means of neutrino
emission from nucleons in a mechanism known as the direct Urca
process~\cite{Lat01_APJ550,Lei02_NPA707,Yak04_ARAA42,Pag04_APJ155}:
\begin{subequations}
 \begin{eqnarray}
  && n \rightarrow p + e^{-} + \bar\nu_{e}\;, \label{urca1}\\
  && e^{-} + p \rightarrow n + \nu_{e}\;.     \label{urca2}
 \end{eqnarray}
\end{subequations}
This mechanism is not exotic as it only relies on protons, neutrons,
electrons, and muons --- standard constituents of dense matter.
However, it requires a large proton fraction $Y_{p}$ for the momentum
to be conserved in the above reactions. As a large proton fraction
requires a stiff symmetry energy, it is interesting to determine if
the newly proposed EOS is able to support such a large proton
fraction. Note that in order for the direct Urca process to operate,
the proton fraction must exceed $Y_{p}\!=\!0.111$ for the low-density
(muonless) case, and $Y_{p}\!=\!0.148$ for the high-density case (with
equal number of electrons and muons). In Table~\ref{Table3} we list
the threshold density ($\rho_{_{\rm Urca}}$) and minimum mass ($M_{\rm
Urca}$) required for the onset of the direct Urca process.  We note
that in spite of its softer symmetry energy, FSUGold predicts that a
$1.4$ solar-mass neutron star may cool down by the direct Urca
process. For completeness, the mass fraction that supports enhanced 
cooling in such a neutron star is listed as $\Delta M_{\rm Urca}$.

In conclusion:

1. A new accurately calibrated relativistic model (``FSUGold'') has
   been fitted to the binding energies and charge radii of a variety
   of magic nuclei. The model is as successful as the NL3 set --- used
   here as a successful paradigm --- in reproducing the ground-state
   properties of a variety of nuclei. Symmetric nuclear
   matter saturates at a Fermi momentum of $k_{\rm F}\!=\!1.30~{\rm
   fm}^{-1}$ (corresponding to a baryon density of $0.15~{\rm
   fm}^{-3}$) with a binding energy per nucleon of
   $B/A\!=\!-16.30$~MeV.  Yet relative to NL3, FSUGold contains two
   additional parameters, whose main virtue is the softening of both
   the EOS of symmetric matter and the symmetry energy.

2. While the need for the two additional parameters is not manifest in
   the description of ground-state properties, it becomes essential
   for reproducing a few nuclear collective modes. Specifically, the 
   breathing mode in ${}^{90}$Zr is sensitive to the softening of 
   symmetric matter, the isovector giant dipole resonance in 
   ${}^{208}$Pb to the softening of the symmetry energy, and the 
   breathing mode in ${}^{208}$Pb to both. Incorporating these
   additional constraints yields a nuclear-matter incompressibility 
   of $K\!=\!230$~MeV and a neutron skin thickness in ${}^{208}$Pb 
   of $R_{n}-R_{p}\!=\!0.21$~fm.

3. While the description of the various collective modes imposes
   additional constraints on the EOS at densities slightly above and
   below saturation density, the high-density component of the EOS
   remains largely unconstrained. We made no attempts to constrain the
   EOS at the supranuclear densities of relevance to neutron-star
   physics. Rather, we simply explored the consequences of the new
   parametrization on a variety of neutron star observables, namely,
   masses, radii, and composition. In particular, we found a limiting
   mass of $M_{\rm max}\!=\!1.72 M_{\odot}$, a radius of
   $R\!=\!12.66$~km for a canonical $M\!=\!1.4 M_{\odot}$ neutron
   star, and no direct Urca cooling in neutrons stars with masses
   below $M\!=\!1.3 M_{\odot}$.  While the consequences of these
   results will be fully explored in a forthcoming publication, it is
   interesting to note that recent observations of pulsar-white dwarf
   binaries at the Arecibo observatory may place important constraints
   on our EOS.  Indeed, Nice, Splaver, and Stairs have inferred a
   pulsar mass for PSRJ0751+1807 of $M\!=\!2.1^{+0.4}_{-0.5}
   M_{\odot}$ at a 95\% confidence level~\cite{Nic04_IAU}. If
   this observation could be refined, not only would it rule out the
   high-density behavior of this (and many other) EOS, but it could 
   provide us with a precious boost in our quest for the equation of 
   state.

\smallskip
We acknowledge many useful discussions with C.J. Horowitz 
and D. Page. This work was supported in part by DOE grant 
DE-FG05-92ER40750.

\bibliography{ReferencesJP}

\end{document}